\documentclass[conference]{IEEEtran}
\IEEEoverridecommandlockouts
\pdfoutput=1
\usepackage{cite}
\usepackage{amsmath,amssymb,amsfonts}
\usepackage{algorithmic}
\usepackage{graphicx}
\usepackage{tikz}
\usepackage[nolist]{acronym}
\usepackage{tabu}
\usepackage{multirow}
\usepackage{multicol}
\usepackage{listings}
\newif\ifarxiv
\arxivtrue 

\newif\ifopen
\newif\iffinal

\ifarxiv
\opentrue
\finaltrue
\fi

\makeatletter
\def\lst@makecaption{%
  \def\@captype{table}%
  \@makecaption
}
\makeatother
\usepackage{booktabs}
\usetikzlibrary{arrows.meta}
\tikzset{%
  >={Latex[width=2mm,length=2mm]},
            base/.style = {rectangle, rounded corners, draw=black,
                           minimum width=4cm, minimum height=1cm,
                           text centered, font=\sffamily},
}
\usepackage{textcomp}
\usepackage[colorinlistoftodos,prependcaption]{todonotes}
\usepackage{regexpatch}
\makeatletter
\xpatchcmd{\@todo}{\setkeys{todonotes}{#1}}{\setkeys{todonotes}{inline,#1}}{}{}
\makeatother
\usepackage{xcolor}
\usepackage[binary-units, per-mode=symbol]{siunitx}
\def\BibTeX{{\rm B\kern-.05em{\sc i\kern-.025em b}\kern-.08em
    T\kern-.1667em\lower.7ex\hbox{E}\kern-.125emX}}

\iffinal
\newcommand{\colorrevi}{black}
\else
\newcommand{\colorrevi}{blue}
\fi
\newcommand{\rev}[1]{{\textcolor{\colorrevi}{#1}}}

\begin{document}

\begin{acronym}
\acro{ALM}{Adaptive Logic Modules}
\acro{BRAM}{Block RAM}
\acro{CPU}{Central Processing Unit}
\acro{DDR}[DDR-SDRAM]{Double Data Rate Synchronous Dynamic Random Access Memory}
\acro{DSP}{Digital Signal Processor}
\acro{FF}{Flop-Flop}
\acro{HBM2}{High Bandwidth Memory 2}
\acro{PSN}[PSN]{Packet Switched Network}
\acro{CSN}{Circuit Switched Network}
\acro{MPI}{Message Passing Interface}
\acro{LUT}{Lookup Table}
\acro{FPGA}{Field Programmable Gate Array}
\acro{HPC}{High Performance Computing}
\acro{HPL}{High Performance LINPACK}
\acro{HLS}{High Level Synthesis}
\acro{HPCC}{HPC Challenge}
\acro{OpenCL}{Open Computing Language}
\acro{GPU}{Graphics Processing Unit}
\acro{FLOPS}{Floating Point Operations per Second}
\acro{SDK}{Software Development Kit}
\acro{BSP}{Board Support Package}
\acro{LSU}{Load Store Unit}
\acro{SVM}{Shared Virtual Memory}
\acro{FLOP}{Floating Point Operation}
\acro{GUPS}{Giga Updates Per Second}
\end{acronym}

\title{\ifthenelse{\NOT \boolean{final}}{\rev{Revision:} }{}Evaluating FPGA Accelerator Performance with a Parameterized OpenCL Adaptation of the HPCChallenge Benchmark Suite}

\ifopen

\author{\IEEEauthorblockN{Marius Meyer\IEEEauthorrefmark{1},
Tobias Kenter\IEEEauthorrefmark{2} and Christian Plessl\IEEEauthorrefmark{3}}
\IEEEauthorblockA{Department of Computer Science and Paderborn Center for Parallel Computing (PC$^2$)\\
Paderborn University, Paderborn, Germany\\
Email: \IEEEauthorrefmark{1}marius.meyer@uni-paderborn.de,
\IEEEauthorrefmark{2}tobias.kenter@uni-paderborn.de,
\IEEEauthorrefmark{3}christian.plessl@uni-paderborn.de}}

\else
\author{\IEEEauthorblockN{Anonymous Author(s)}}
\fi

\maketitle

\begin{abstract}

FPGAs have found increasing adoption in data center applications since a new generation of high-level tools have become available which noticeably reduce development time for FPGA accelerators and still provide high-quality results. There is, however, no high-level benchmark suite available, which specifically enables a comparison of FPGA architectures, programming tools, and libraries for HPC applications.

To fill this gap, we have developed an OpenCL-based open-source implementation of the HPCC benchmark suite for Xilinx and Intel FPGAs. This benchmark can serve to analyze the current capabilities of FPGA devices, cards, and development tool flows, track progress over time, and point out specific difficulties for FPGA acceleration in the HPC domain. Additionally, the benchmark documents proven performance optimization patterns. We will continue optimizing and porting the benchmark for new generations of FPGAs and design tools and encourage active participation to create a valuable tool for the community.

\end{abstract}

\acresetall


\acresetall

\acused{CPU}

\section{Introduction}
In \ac{HPC}, benchmarks are an important tool for performance comparison across systems.
They are designed to stress important system properties or generate workloads that are similar to relevant applications for the user.
Especially in acquisition planning they can be used to define the desired performance of the acquired system before it is built.
Since it is a challenging task to select a set of benchmarks to cover all relevant device properties, benchmark suites can help by providing a pre-defined mix of applications and inputs, for example SPEC CPU~\cite{SPEC-CPU} and \ac{HPCC}~\cite{HPCCIntroduction}.

There is an ongoing trend towards heterogeneity in \ac{HPC}, complementing \acp{CPU} by accelerators, as indicated by the Top 500 list~\cite{top500}.
From the top 10 systems in the list, seven are equipped with different types of accelerators.
Nevertheless, to get the best matching accelerator for a new system, a tool is needed to measure and compare the performance across accelerators.
For well-established accelerator architectures like \acp{GPU}, there are already standardized benchmarks like SPEC ACCEL \cite{SPECACCEL}. For \acp{FPGA}, that are just emerging as accelerator architecture for data centers and HPC, existing benchmarks
do not focus on \ac{HPC} and miss to measure highly relevant device properties.

Similar to the compiler for \ac{CPU} applications, the \ac{HLS} framework consisting of \ac{SDK} and \ac{BSP} takes a very important role to achieve performance on an \ac{FPGA}.
The framework translates the accelerator code (denoted as \emph{kernel}), most commonly from \ac{OpenCL}, to intermediate languages, organizes the communication with the underlying \ac{BSP}, performs optimizations and synthesizes the code to create executable \ac{FPGA} configurations (bitstreams).
Hence, the \ac{HLS} framework has a big impact on the used \ac{FPGA} resources and the maximum kernel frequency, which might vary depending on the kernel design.
An \ac{HPC} benchmark suite for \acp{FPGA} should capture this impact and, for comparisons, must not be limited to a single \ac{HLS} framework.

One of the core aspects of \ac{HPC} is communication. Some \ac{FPGA} cards offer a new approach for scaling with their support for direct communication to other \ac{FPGA} cards without involving the host CPU. 
Such technology is already used in first applications~\cite{MLNetwork,Sano-Multi-FPGA-Stencil} and research has started to explore the best abstractions and programming models for inter-FPGA communication~\cite{SMI, FPGAEthernet}. 
Thus, communication between FPGAs out of an \ac{HLS} framework is another essential characteristic that a \ac{FPGA} benchmark suite targeting \ac{HPC} should consider.

In this paper, we propose \emph{HPCC FPGA}, an \ac{FPGA} \ac{OpenCL} benchmark suite for \ac{HPC} using the applications of the \ac{HPCC} benchmark suite. The motivation for choosing HPCC is that it is well-established for CPUs and covers a small set of applications that evaluate important memory access and computing patterns that are frequently used in HPC applications. 
Further, the benchmark also characterizes the HPC system's network bandwidth allowing to extrapolate to the performance of parallel applications.


Specifically, we make the following contributions in this paper: 
%
\begin{enumerate}

    \item We provide FPGA-adapted \ac{OpenCL} kernel implementations along with corresponding host code for setup and measurements for all \ac{HPCC} benchmark applications.
    \item We provide configuration options for the \ac{OpenCL} kernels that allow adjustments to resources and architecture of the target \ac{FPGA} and board without the need to change the code manually.
    
    \item We evaluate the execution of these benchmarks on different FPGA families and boards with Intel and Xilinx FPGAs and show the benchmarks can capture relevant device properties.
    
    \item We make all benchmarks and the build system available as open-source on GitHub to encourage community contributions.

\end{enumerate}

The remainder of this paper is organized as follows: In Section~\ref{sec:related-work}, we give an overview of existing \ac{FPGA} benchmark suites.
In Section~\ref{sec:hpcc-benchmark}, we introduce the benchmarks in \ac{HPCC}  \ac{FPGA} in more detail and briefly discuss the contained benchmarks and the configuration options provided for the base runs. In Section~\ref{sec:evaluation} we build the benchmarks for different \ac{FPGA} architectures and evaluate the results to show the potential of the proposed configurable base runs. 
In Section~\ref{sec:discussion}, we evaluate the global memory system of the boards in more detail and give insights into experienced problems and the potential of the benchmarks to describe the performance of \ac{FPGA} boards and the associated frameworks.
Finally, in Section~\ref{sec:conclusion}, we draw conclusions and outline future work.

\section{Related Work}
\label{sec:related-work}
There already exist several benchmark suites for \acp{FPGA} and their \ac{HLS} frameworks.
Most \ac{OpenCL} benchmark suites like Rodinia~\cite{Rodinia}, OpenDwarfs~\cite{OpenDwarfs-first} or SHOC~\cite{SHOC} are originally designed with GPUs in mind.
Although both GPU and \ac{FPGA} can be programmed using \ac{OpenCL}, the design of the compute kernels has to be changed and optimized specifically for \ac{FPGA} to achieve good performance.
In the case of Rodinia this was done~\cite{RodiniaFPGA} for a subset of the benchmark suite with a focus on different optimization patterns for the Intel FPGA (then Altera) SDK for OpenCL.
In contrast, to port OpenDwarfs to FPGAs, Feng~et~al.~\cite{OpenDwarfs-first} employed a research OpenCL synthesis tool that instantiates GPU-like architectures on FPGAs.
With Rosetta~\cite{Rosetta}, there also exists a benchmark suite that was designed targeting \acp{FPGA} using the Xilinx HLS tools from the start.
It focuses on typical FPGA streaming applications from the video processing and machine learning domains. 
The CHO~\cite{CHO} benchmark targets more fundamental FPGA functionality and includes kernels from media processing and cryptography and the low-level generation of floating-point arithmetic through OpenCL, using the Altera SDK for OpenCL.

The mentioned benchmarks often lack possibilities to adjust the benchmarks to the target \ac{FPGA} architecture easily.
Modifications have to be done manually in the kernel code, sometimes many different kernel variants are proposed or the kernels are not optimized at all, making it difficult to compare results for different \acp{FPGA}.
A benchmark suite that takes a different approach is Spector \cite{Spector}.
It makes use of several optimization parameters for every benchmark, which allows modification and optimization of the kernels for a \ac{FPGA} architecture.
The kernel code does not have to be manually changed, and optimization options are restricted by the defined parameters.
Nevertheless, the focus is more on the research of the design space than on performance characterization.

To our best knowledge, there exists no \ac{OpenCL} benchmark suite for \ac{FPGA} with a focus on \ac{HPC} characteristics at the point of writing.
All of the mentioned benchmark suites lack a way to measure the inter-\ac{FPGA} communication capability of recent high-end \acp{FPGA}. In some of the benchmarks, the investigated input sizes are small enough to fit into local memory resources of a single \ac{FPGA}.
Since actual \ac{HPC} applications are highly parallel and require effective communication, an \ac{HPC} focused benchmark must also evaluate the characteristics of the communication network.

\section{HPC Challenge Benchmarks for FPGA}
\label{sec:hpcc-benchmark}
%

The \ac{HPCC} benchmark suite~\cite{HPCCIntroduction} consists of seven benchmarks.
Three of them are synthetic benchmarks that measure the memory performance for successive accesses (STREAM \cite{STREAM}) and random updates (RandomAccess) as well as the effective network bandwidth of a system (b\_eff).
Especially the latter will give important insights for the use in \ac{HPC} because of the inter-\ac{FPGA} communication support in recent devices.
Moreover, it contains four applications of varying complexity:
Matrix transposition (PTRANS), 1D Fast Fourier Transformation (FFT), matrix multiplication (GEMM) and \ac{HPL}.
\rev{All benchmarks except RandomAccess and b\_eff use single-precision floating-point values for calculation since this is most commonly used in \ac{FPGA} designs.
RandomAccess is using \SI{64}{\bit} integers because of the used pseudo-random number scheme and b\_eff \SI{8}{\bit} integers.
Nevertheless, the build system allows to change the used data type for all benchmarks easily.
}
%
%
A central concept of the \ac{HPCC} benchmark suite is to characterize the performance of different memory access patterns of \ac{HPC} applications.
By determining the spatial and temporal dependencies of the memory accesses for another application, the benchmark results can then be used to estimate the performance of this application on the system more precisely.
The achieved performance in \ac{CPU} systems is thus highly dependent on the memory and cache architecture.


When benchmarking \ac{FPGA} boards with this concept, it is crucial to note that even for a given board, the memory hierarchy is not fixed. While the off-chip memory itself and typically parts of the memory controllers are fixed hardware components, local buffers or caches, address generators, pre-fetchers, and data buses inside the FPGA are provided by the \ac{BSP} or generated by the \ac{SDK} depending on the benchmark implementation. Thus, in contrast to the CPU version of HPCC, HPCC FPGA does not only measure hardware properties of a given memory interface but rather also the ability of the tools to optimize the memory hierarchy and compute pipelines of a kernel for the specific pattern to provide good performance of the calculation on an \ac{FPGA}.

Another aspect of the \ac{HPCC} benchmark suite is the distinction between two different runs:
\begin{itemize}
    \item \emph{Base runs} are done with the provided reference implementations.
    \item \emph{Optimized runs} are done with implementations that use architecture-specific optimizations.
\end{itemize}

The goal of HPCC FPGA, as presented in this article, is to provide implementations for the base runs that are reasonably optimized for Intel and Xilinx FPGAs and their SDKs. With current FPGA execution models, such optimizations necessarily include adaptations to the available resources, in particular the number of physical memory interfaces and configurable local memory. Therefore, the base implementation exposes defined configuration parameters for such customization, without pre-empting the full flexibility that is reserved for optimized runs with manual code changes and target architecture- or SDK-specific designs. The presented adjustable parameters for the base runs are a subject of discussion and might be changed with the evolution of the \ac{FPGA} architectures and toolchains.

In Table~\ref{tab:benchmark_mem_patterns} the memory access patterns for the benchmarks contained in \ac{HPCC} are given for CPU and the \ac{FPGA} base implementations proposed in this paper.
For \ac{FPGA}, the memory is further divided into global and local memory representing the DDR RAM or \ac{BRAM} and registers, respectively.
Spatial locality is represented by the \emph{linear} and \emph{blocked} access pattern, while temporal locality is separately indicated with a \emph{T}.
Since it is possible to partially define the memory hierarchy used for the application, the \ac{FPGA} designs attempt to move the strided memory accesses with low spatial locality into the local memory and increase the temporal locality if possible using blocked algorithms.
The global memory again is accessed in a blocked fashion to increase spatial locality and decrease temporal locality.

\begin{table}
    \centering
    \caption{Comparison of the HPCC Benchmarks and their memory access patterns on CPU and FPGA}
    \begin{tabular*}{\linewidth}{p{1.6cm}||p{1.9cm}||p{1.8cm}|p{1.9cm}}
    \toprule
        \multirow{2}{*}{\textbf{Benchmark}} & \multirow{2}{*}{\textbf{CPU}} & \multicolumn{2}{c}{\textbf{FPGA}} \\ 
         && \textbf{Global Memory} & \textbf{Local Memory} \\
         \midrule
STREAM & linear & linear & linear \\
RandomAccess & random & random & linear \\
PTRANS & strided, linear & blocked, linear & strided, linear \\
FFT & strided, T & linear & strided, T\\
GEMM & strided, linear, T & blocked, linear & strided, linear, T\\
LINPACK & strided, linear, T & blocked,linear & strided, linear, T \\
         \bottomrule
         \multicolumn{4}{l}{T = temporal locality}
    \end{tabular*}
    \label{tab:benchmark_mem_patterns}
\end{table}

The benchmark suite is open source and publicly available on GitHub.
\ifopen
\footnote{https://github.com/pc2/HPCC\_FPGA}
\else
\footnote{Link will be made available here after review}
\fi



\subsection{Common Build Setup}

The HPCC FPGA benchmark suite is set up to create one host binary and FPGA bitstream per benchmark, with source code structure, build process, and benchmark execution very similar for all benchmarks. For usability, HPCC FPGA adopts the following approaches:

\begin{itemize}
    \item Usage of the same build system for all benchmarks (CMake) offers a unified user experience during the build process and for the modification of the configuration parameters.
    \item Integrated tests allow checking functional correctness of the configuration using emulation before actual synthesis.
\end{itemize}

A first set of parameters (Table~\ref{tab:common_parameter_desc}) is exposed to select the target FPGA board (\texttt{FPGA\_BOARD\_NAME}) and provide default parameters for the execution of the OpenCL host binary. The latter values can still be changed at execution time, for example, to test several identical devices in one server sequentially.
Additional compiler arguments can be used with the Intel and Xilinx compiler to control the synthesis or trigger special optimizations (Table~\ref{tab:common_parameter_strema_ra_desc}).
While the Intel FPGA SDK for OpenCL options are provided as \texttt{AOC\_FLAGS}, the Xilinx Vitis toolchain expects parameters or configuration files at several stages. They are used, for example, to create multiple kernel copies for the STREAM and RandomAccess benchmarks.
Finally, for each benchmark, specific parameters are exposed to make good use of the board's memory interfaces and local memory of the FPGA. These parameters are summarized in Tables~\ref{tab:stream_parameter_desc}--\ref{tab:linpack_parameter_desc} with the benchmarks described in the following subsections.

\begin{table}
    \centering
    \caption{The build parameters to select the target device at compile time and default values at execution time}
    \begin{tabular*}{\linewidth}{p{3cm}p{4.5cm}}
    \toprule
        \textbf{Parameter} &
        \textbf{Description} \\
         \midrule
\texttt{FPGA\_BOARD\_NAME}  & Name of the target board \\
\midrule
\texttt{DEFAULT\_DEVICE} & Index of the default device \\
\midrule
\texttt{DEFAULT\_PLATFORM} & Index of the default platform \\
\midrule
\texttt{DEFAULT\_REPETITIONS} & Number of times the kernel will be executed\\
         \bottomrule
    \end{tabular*}
    \label{tab:common_parameter_desc}
\end{table}


\begin{table}
    \centering
    \caption{Tool-specific optimization flags as exposed through the CMake build system of HPCC FPGA}
    \begin{tabular*}{\linewidth}{p{3.8cm}p{4cm}}
    \toprule
        \textbf{Parameter} &
        \textbf{Description} \\
         \midrule
\texttt{AOC\_FLAGS} & Additional compiler flags that are used for kernel compilation with the Intel toolchain\\
\midrule
\texttt{XILINX\_COMPILE\_FLAGS} & Additional compiler flags that are used for kernel compilation with the Xilinx Vitis toolchain\\
\midrule
\texttt{XILINX\_COMPILE\_SETTINGS} & Path to the settings file that contains additional compile options for the Xilinx Vitis toolchain\\
\midrule
\texttt{XILINX\_GENERATE\_LINK \_SETTINGS} & Boolean that can be set to trigger link setting generation using a file template for Xilinx Vitis\\
\midrule
\texttt{XILINX\_LINK\_SETTINGS} & Path to the settings file that contains the link options for the Xilinx Vitis toolchainn\\
         \bottomrule
    \end{tabular*}
    \label{tab:common_parameter_strema_ra_desc}
\end{table}

\subsection{STREAM benchmark}

The goal of the STREAM benchmark is to measure the sustainable memory bandwidth of a device using four different vector operations: \emph{Copy}, \emph{Scale}, \emph{Add} and \emph{Triad}.
All kernels are taken from the STREAM benchmark\footnote{https://www.cs.virginia.edu/stream/} v5.10 and thus slightly differ from the kernels proposed in the \ac{HPCC} article~\cite{HPCCIntroduction}.
The benchmark will sequentially execute the operations given in Table~\ref{tab:stream_kernels}.
\emph{PCI Read} and \emph{PCI Write} are no \ac{OpenCL} kernels but represent the read and write of the arrays to the device memory.
The benchmark will output the maximum, average, and minimum times measured for all of these operations.
The minimum time will also be used to calculate the memory bandwidth for the operation.

\begin{table}
    \centering
    \caption{The STREAM benchmark kernels that are calculating on three arrays $A$, $B$ and $C$ and the PCIe transfers that are measured in the order they are executed}
    \begin{tabular}{ll}
    \toprule
        \textbf{Name} & \textbf{Kernel Logic} \\
         \midrule
         PCI Write & write arrays to device \\
         Copy & $C[i] = A[i]$ \\
         Scale & $B[i] = j * C[i]$ \\
         Add & $C[i] = A[i] + B[i]$\\
         Triad & $A[i] = j * C[i] + B[i]$ \\
         PCI Read & read arrays from device\\
         \bottomrule
    \end{tabular}
    \label{tab:stream_kernels}
\end{table}

The arrays $A$, $B$, and $C$ are initialized with a constant value over the whole array.
This allows us to validate the result by only recalculating the operations with scalar values.
The error is calculated for every value in the arrays and must be below the machine epsilon $\epsilon < ||d - d'||$ to pass the validation.

A simplified version of the code is given in Listing~\ref{lst:stream}.
The \ac{OpenCL} kernel of the benchmark combines all four described compute kernels in a single function.
Since on \acp{FPGA} the source code is translated to spatial structures that take up resources on the device, a single combined kernel allows for best reuse of those resources.
The computation is split into blocks of a fixed length, which makes it necessary that the arrays have the length of a multiple of the block size.
In the first inner loop, the input values of the first input array \texttt{in1} are loaded into a buffer located in the local memory of the \ac{FPGA}.
While they are loaded, they are multiplied with a scaling factor \texttt{scalar}.
This allows the execution of the \emph{Copy} and \emph{Scale} operation.
In the case of \emph{Copy}, the scaling factor is set to 1.0.
In the second loop, the second input \texttt{in2} is only added to the buffer, if a flag is set.
Together with the first loop, this makes it possible to recreate the behavior of the \emph{Add} and \emph{Triad} operation.
In the third loop, the content of the buffer is stored in the output array \texttt{out} located in global memory.
\begin{lstlisting}[float,caption=Simplified kernel logic of the STREAM benchmark,
                    label=lst:stream]
for(uint i = 0; i < array_size;
                    i += BUFFER_SIZE) {
    float buffer[BUFFER_SIZE];
    for (uint k = 0;k < BUFFER_SIZE;
                                 k++) {
        buffer[k] = scalar * in1[i + k];
    }
    if (second_input) {
        for (uint k = 0;k < BUFFER_SIZE;
                                 k++) {
            buffer[k] += in2[i + k];
        }
    }
    for (uint k = 0;k < BUFFER_SIZE;
                                k++) {
        out[i + k] = buffer1[k];
               
    }
}
\end{lstlisting}

At build time, the parameters given in Figure~\ref{tab:stream_parameter_desc} can be modified to generate a the base run kernel.
\texttt{DATA\_TYPE} and \texttt{VECTOR\_COUNT} can be used to define the data type used within the kernel.
If \texttt{VECTOR\_COUNT} is greater than 1, \ac{OpenCL} vector types of the given length are used.
Moreover, it is possible to adjust the resource usage of the kernel by specifying the size of the local memory buffer and the unrolling factor of the three loops of the kernel.
The intended use case of the \ac{OpenCL} kernel is to replicate the kernel for every available memory bank.
The host code will split and distribute the arrays equally to all memory banks, so every kernel will just have a fraction of the whole arrays to update.


\begin{table}
    \centering
    \caption{The configuration options that are exposed to the user to modify the kernel generation of the STREAM benchmark}
    \begin{tabular}{p{3cm}p{4.5cm}}
    \toprule
        \textbf{Parameter} &
        \textbf{Description} \\
         \midrule
\texttt{DATA\_TYPE} &  Data type used for host and device code \\
\midrule
\texttt{VECTOR\_COUNT} &  If $>1$ OpenCL vector types of the given size are used in the device code \\
\midrule
\texttt{GLOBAL\_MEM\_UNROLL}  & Loop unrolling factor for the inner loops in the device code \\
\midrule
\texttt{NUM\_REPLICATIONS} &  Replicates the kernels the given number of times \\
\midrule
\texttt{DEVICE\_BUFFER\_SIZE} &  Number of values that are stored in the local memory in the single kernel approach \\
         \bottomrule
    \end{tabular}
    \label{tab:stream_parameter_desc}
\end{table}

\subsection{RandomAccess benchmark}

The random access benchmark measures the performance for non-consecutive memory accesses expressed in the performance metric \ac{GUPS}.
It updates values in an data array $d \in \Bbb Z^n$ such that $d_{i} = d_{i} \oplus a$ where $a \in \Bbb Z$ is a value from a pseudo random sequence.
$n$ is defined to be a power of two. 
Since only operations in a finite field are used to update the values, the correctness of the updates can easily be checked by executing a reference implementation on the host side on the resulting data.
The incorrect items are counted, and the error percentage is calculated with $\frac{error}{n}*100$.
An error of $<1\%$ has to be accomplished to pass the validation. Hence, update errors caused by concurrent data accesses are tolerated to some degree.
Similar to the STREAM implementation the benchmark allows us to replicate the kernel for every memory bank.
\rev{The benchmark allows to store subsequent reads and writes from the global memory in a local memory buffer.
On the one hand, this allows us to partially hide the latency of a single memory access and increase the performance of the benchmark.
On the other hand, this mechanism will lead to errors if the same memory address is loaded multiple times into the buffer.
Since the benchmark allows a small amount of errors, changing the buffer size can be used as a trade-off between benchmark error and performance.
}
The local memory buffer size and the number of replications can be configured with build parameters that are given in Table~\ref{tab:ra_parameter_desc}.


    

\begin{table}
    \centering
    \caption{The configuration options that are exposed to the user to modify the kernel generation of the Random Access benchmark}
    \begin{tabular}{p{3cm}p{4.5cm}}
    \toprule
        \textbf{Parameter} &
        \textbf{Description} \\
         \midrule
\texttt{NUM\_REPLICATIONS} &  Replicates the kernels the given number of times \\
\midrule
\texttt{DEVICE\_BUFFER\_SIZE} &  Number of values that are stored in the local memory in the single kernel approach \\
         \bottomrule
    \end{tabular}
    \label{tab:ra_parameter_desc}
\end{table}

\subsection{Effective Bandwidth Benchmark (b\_eff)}

\emph{b\_eff} is a benchmark that measures the effective network bandwidth.
It is originally designed for the use with \ac{MPI} and thus is not straight forward to translate the same functionality to \ac{OpenCL} and the I/O channels of the \ac{FPGA}.
For inter-\ac{FPGA} communication, there exists no such default communication paradigm and also, the network layer is a matter of current research \cite{SMI, FPGAEthernet}.
Thus, the current implementation of the network benchmark does not consider routing overheads and different network topologies as it is done in the original benchmark.
Instead, only a single ring topology containing all \acp{FPGA} is used to execute the benchmark.

The benchmark contains two kernels \emph{send} and \emph{recv} that will alternately send an receive messages over the I/O channels.
The \emph{send} kernel will first send a message of a given size and then receive the message, whereas the \emph{recv} kernel will first receive and then send a message.
This process is repeated for an adjustable number of iterations to reduce the kernel start overhead in the measurements.
The benchmark execution can thus be scaled to an arbitrary number of \acp{FPGA}.

The benchmark executes the kernels for 21 different message sizes $L={2^0,2^1,\dots,2^{20}}$.
Based on that, the effective network bandwidth is calculated with Equation~\ref{eq:effective_bandwidth}.

\begin{equation}
    b_{\mathrm{eff}} = \frac{\sum_L b_L}{21}
    \label{eq:effective_bandwidth}
\end{equation}
with $b_L$ being the measured bandwidth for message size $L$.


\begin{table}
    \centering
    \caption{The configuration options that are exposed to the user to modify the kernel generation of the b\_eff benchmark}
    \begin{tabular}{p{3cm}p{4.5cm}}
    \toprule
        \textbf{Parameter} &
        \textbf{Description} \\
         \midrule
\texttt{CHANNEL\_WIDTH}  & Channel width in Bytes\\
         \bottomrule
    \end{tabular}
    \label{tab:beff_parameter_desc}
\end{table}

The modifiable parameters for b\_eff are given in Table~\ref{tab:beff_parameter_desc}.
So far, only the width of a single channel can be varied.
External channels are currently very specific to the used device because of the missing layer of abstraction like it would be done in MPI, so manual modification is allowed for this kernel file.
The host code of the benchmark uses \ac{MPI} for communication and to collect the runtime measurements of the kernels.

\subsection{Matrix Transposition Benchmark (PTRANS)}

\begin{table}
    \centering
    \caption{The configuration options that are exposed to the user to modify the kernel generation of the PTRANS benchmark}
    \begin{tabular}{p{3cm}p{4.5cm}}
    \toprule
        \textbf{Parameter} &
        \textbf{Description} \\
         \midrule
\texttt{BLOCK\_SIZE}  & Size of the symmetric matrix block that will be stored in local memory. Should have a sufficient size to allow full utilization of global memory read and write bursts.\\
\midrule
\texttt{GLOBAL\_MEM\_UNROLL}  & Number of times the loops loading and storing to global memory have to be unrolled to create \acp{LSU} with the same width than the memory interface\\
         \bottomrule
    \end{tabular}
    \label{tab:ptrans_parameter_desc}
\end{table}

The PTRANS benchmark works on square matrices $A,B,C \in \Bbb R^{n\times n}$ and calculates $C = A^T + B$, i.e. it transposes a matrix $A$ and adds $B$ to the result.
The number of \acp{FLOP} is $n^2$ for $A,B,C \int \Bbb R^{n \times n}$ for this calculation.
The kernel execution time is measured with the host code and used to calculate the final performance metric \ac{FLOPS}.
The execution is validated by calculating the residual $\frac{||C-C'||}{\epsilon n}$ where $C'$ is the result of a reference implementation, $\epsilon$ is the machine epsilon and $n$ dimension of the matrices.

\subsection{Fast Fourier Transformation Benchmark (FFT)}

\begin{table}
    \centering
    \caption{The configuration options that are exposed to the user to modify the kernel generation of the FFT benchmark}
    \begin{tabular}{p{3cm}p{4.5cm}}
    \toprule
        \textbf{Parameter} &
        \textbf{Description} \\
         \midrule
\texttt{LOG\_FFT\_SIZE}  & Logarithm of the FFT size that should be calculated. It should be as big as possible to make use of a much resources as possible and to utilize the global memory bursts. The size is limited by the implementation to 12.\\
         \bottomrule
    \end{tabular}
    \label{tab:fft_parameter_desc}
\end{table}

In the FFT benchmark a 1d FFT is calculated with a size of up to $2^{12}$ for single precision complex numbers.
The benchmark is based on a reference implementation for the Intel OpenCL FPGA SDK included in version 19.4.0.
Since the calculation of a single FFT would lead to very short execution times that lead to high measurement errors, batch processing is used to increase the overall execution time.
This also allows better utilization of the kernel pipeline.
The number of \acp{FLOP} for this calculation is defined to be $5*n*ld(n)$ for an FFT of dimension $n$.
The result of the calculation is checked by calculating the residual $\frac{||d - d'||}{\epsilon ld(n)}$ where $\epsilon$ is the machine epsilon, $d'$ the result from the reference implementation and $n$ the FFT size.

\subsection{Matrix Multiplication Benchmark (GEMM)}

\begin{table}
    \centering
    \caption{The configuration options that are exposed to the user to modify the kernel generation of the DGEMM benchmark}
    \begin{tabular}{p{3cm}p{4.5cm}}
    \toprule
        \textbf{Parameter} &
        \textbf{Description} \\
         \midrule
\texttt{BLOCK\_SIZE}  & Size of the symmetric matrix block that will be stored in local memory. Should have a sufficient size to allow full utilization of global memory read and write bursts.\\
\midrule
\texttt{GEMM\_SIZE}  & Size of the symmetric matrix block that will be stored in registers. This will affect the amount of \acp{DSP} used in the implementation.\\
\midrule
\texttt{GLOBAL\_MEM\_UNROLL}  & Number of times the loops loading and storing to global memory have to be unrolled to create \acp{LSU} with the same width than the memory interface\\
         \bottomrule
    \end{tabular}
    \label{tab:dgemm_parameter_desc}
\end{table}

The GEMM benchmark implements a matrix-matrix multiplication similar to the GEMM routines in the BLAS library.
It calculates $C = \alpha * A * B + \beta * C$ where $A,B,C \in \Bbb R^{n \times n}$ and $\alpha, \beta \in \Bbb R$.
The number of \acp{FLOP} for the performance calculation is defined to be $2 * n^3$.
The result is verified by calculating the residual $\frac{||C - C'||}{\epsilon n ||C||_F}$ where $\epsilon$ is the machine epsilon and $C'$ the result of the reference implementation.
The implementation is based on a matrix multiplication design for Intel Stratix 10 proposed by Gorlani et al. \cite{matrix_mul} and simplified to make it compatible with a broader range of devices.



\subsection{High-Performance Linpack Benchmark (HPL)}

\begin{table}
    \centering
    \caption{The configuration options that are exposed to the user to modify the kernel generation of the LINPACK benchmark}
    \begin{tabular}{p{3cm}p{4.5cm}}
    \toprule
        \textbf{Parameter} &
        \textbf{Description} \\
         \midrule
\texttt{LOCAL\_MEM\_BLOCK\_LOG}  & Size of the symmetric matrix block that will be stored in local memory. Should have a sufficient size to allow full utilization of global memory read and write bursts.\\
\midrule
\texttt{REGISTER\_BLOCK\_LOG}  & Size of the symmetric matrix block that will be stored in registers. This will affect the amount of \acp{DSP} used in the implementation.\\
\midrule
\texttt{GLOBAL\_MEM\_UNROLL}  & Number of times the loops loading and storing to global memory have to be unrolled to create \acp{LSU} with the same width than the memory interface\\
         \bottomrule
    \end{tabular}
    \label{tab:linpack_parameter_desc}
\end{table}

The \ac{HPL} benchmark solves a linear system of equations of order $n$: $Ax=b$ where $A \in \Bbb R^{n \times n}$ and $x,b, \in \Bbb R^n$.
The original benchmark first calculates the LU factorization with row-wise partial pivoting of the matrix $A$ such that $P[A,b]=[[L,U],y]$
In a second step, the linear equation system can be solved by first solving $Ly=b$ and then $Ux=y$.
The \acp{FLOP} for the factorization are $\frac{2}{3}n^3-\frac{1}{2}n^2$ and for solving the linear equations $2n^2$.

Currently, the benchmark is only partially implemented on \ac{FPGA}.
The kernel implementation is based on a blocked approach proposed in \cite{fpga_lu_factorization}.
A LU factorization kernel \rev{ that corresponds to the LINPACK \emph{gefa} routine} is implemented \rev{on the FPGA. 
It uses block-wise partial pivoting instead of partial pivoting over the whole matrix. This design decision was made to reduce the complexity of the kernel}.
The equation system is solved on the CPU and not taken into account for the kernel performance.
The resulting performance metric is \ac{FLOPS} and the result is validated by a reference implementation that checks the residual $\frac{||Ax - b||}{\epsilon ||A|| n}$.


\section{Benchmark Execution and Evaluation}
\label{sec:evaluation}

In the following, we synthesize and execute all benchmarks of the suite, collect first benchmark results for the proposed base implementations, and evaluate the results to performance models.

\subsection{Evaluation Environment and Configuration}

The benchmarks were synthesized and tested on a cluster containing multiple \emph{Nallatech 520N} cards and research systems with \emph{Intel PAC D5005} and \emph{Xilinx Alveo U280} FPGA boards.
The Nallatech 520N boards are connected to the host node via x8 PCIe 3.0 and are equipped with Intel Stratix 10 GX2800 with access to four banks of DDR4 SDRAM x 72 bit with 8 GB per bank and a transfer rate of 2400MT/s.
Moreover, up to 32 \acp{FPGA} are connected within an inter-\ac{FPGA} \ac{CSN}.
Every \ac{FPGA} has four \SI{40}{\giga\bit\per\second} full-duplex links that are connected through a CALIENT S320 Optical Circuit Switch which allows the creation of arbitrary network topologies.
In the \ac{OpenCL} code, these streaming board-to-board connections are exposed as OpenCL pipes or channels.
With the \ac{BSP} version 19.2.0 and \ac{SDK} version 19.4.0 the most recent versions available at the time of writing are used for synthesis and execution.

The Intel PAC D5005 board hosts a Stratix 10 SX2800 \ac{FPGA} and is connected to the host with x16 PCIe 3.0. \rev{The board contains a DDR4 memory infrastructure similar to the other boards, which is however, not used for these experiments. Instead, a \rev{reference design \ac{BSP} (18.1.2\_svm)} was used, that offers direct access to the host's memory using \ac{SVM} by building upon the the Intel Acceleration Stack (IAS)\cite{IAS} version 1.2. The OpenCL kernel compilation was performed with the \ac{SDK} version 19.4.0.}
The host node for the Nallatech 520N and Intel PAC D5005 is a two-socket system equipped with Intel Xeon Gold 6148 \acp{CPU} and \SI{192}{\giga\byte} of DDR4-2666 main memory.
In the case of the Nallatech 520N, two \acp{FPGA} are connected to a single node.
Except for the \emph{b\_eff} benchmark, only one of the \acp{FPGA} is used to execute the benchmark.
The host code is compiled with GCC 8.3.0 and CMake 3.15.3 is used for the configuration and build of the benchmarks.

The Alveo U280 boards are equipped with the XCU280 \ac{FPGA}.
The board is connected to an Intel Xeon Gold 6234 \ac{CPU} over x8 PCIe 4.0 and equipped with two banks of DDR4 SDRAM x 72\,bit with 16\,GB per bank and a transfer rate of 2400\,MT/s.
Moreover the \ac{FPGA} is equipped with \SI{8}{\giga\byte} of \ac{HBM2} split over 32 banks.
The Xilinx Vitis \ac{SDK} is used in version 2019.2 and the shell version is 2019.2.3.
The host code is compiled with GCC 7.4.0 and CMake 3.3.2 is used for the configuration and build of the benchmarks.
The \ac{CPU} has access to \SI{108}{\giga\byte} of main memory.

\begin{table*}
    \centering
    \caption{Synthesis configurations of all benchmarks}
    \begin{tabular}{p{1.2cm}p{2.5cm}>{\raggedleft\arraybackslash}p{0.8cm}>{\raggedleft\arraybackslash}p{0.8cm}>{\raggedleft\arraybackslash}p{0.8cm}>{\raggedleft\arraybackslash}p{0.8cm}|p{1.2cm}p{2.5cm}>{\raggedleft\arraybackslash}p{0.8cm}>{\raggedleft\arraybackslash}p{0.8cm}>{\raggedleft\arraybackslash}p{0.8cm}}
    \toprule
        \textbf{Benchmark} &
        \textbf{Parameter} &
        \textbf{520N} &
        \textbf{U280 DDR} &
        \textbf{U280 HBM2} &
        \textbf{PAC SVM} &         
        \textbf{Benchmark} &
        \textbf{Parameter} &
        \textbf{520N}&
        \rev{\textbf{U280 DDR}}&
        \rev{\textbf{PAC SVM}} \\
         \midrule
  \multirow{5}{*}{STREAM}       &\texttt{DATA\_TYPE} & float & float & float & float & \multirow{2}{*}{PTRANS}&\texttt{BLOCK\_SIZE}  & 512 & \rev{512} & \rev{512}\\
&\texttt{VECTOR\_COUNT} &   16 & 16 & 16 & 16 & & \texttt{GLOBAL\_MEM\_UNROLL}  & 16 & \rev{16} & \rev{16}\\
&\texttt{GLOBAL\_MEM\_UNROLL}  & 1 & 1 & 1 & 1 & & &\\
&\texttt{NUM\_REPLICATIONS} &   4 & 2 & 32 & 1&\multirow{3}{*}{GEMM}& \texttt{BLOCK\_SIZE}  & 256 & \rev{256} & \rev{256} \\
&\texttt{DEVICE\_BUFFER\_SIZE} &   4,096 & 16,384 & 2,048 & 1&&\texttt{GEMM\_SIZE}  & 8 & \rev{8} & \rev{8} \\
&&   & &  & &&\texttt{GLOBAL\_MEM\_UNROLL}  & 16 & \rev{16} & \rev{16}\\ 

    Random  &       
    \texttt{NUM\_REPLICATIONS} & 4 & 2 & 32 & 1 & & & \\
    Access &\texttt{DEVICE\_BUFFER\_SIZE} & 1 & 1,024 & 1,024 & 1,024 &\multirow{3}{*}{LINPACK} & \texttt{LOCAL\_MEM\_BLOCK\_LOG}  & 5 && \rev{5}\\
&&  &  & &  &&\texttt{REGISTER\_BLOCK\_LOG}  & 3 && 3\\
b\_eff & \texttt{CHANNEL\_WIDTH}  & 32 & & & & &\texttt{GLOBAL\_MEM\_UNROLL}  & 16 && \rev{16}\\ 
&&  &  & &  && & \\
FFT & \texttt{LOG\_FFT\_SIZE}  & 12 & & &\rev{12} & & &\\
         \bottomrule
    \end{tabular}
    \label{tab:used_parameters}
\end{table*}

\begin{table*}
    \centering
    \caption{Resource usage of the synthesized STREAM and RandomAccess benchmark kernels}
    \begin{tabular}{p{1.4cm}p{1.4cm}>{\raggedleft\arraybackslash}p{1.1cm}@{\hskip 3pt}p{1.1cm}>{\raggedleft\arraybackslash}p{1.1cm}@{\hskip 3pt}p{1.1cm}>{\raggedleft\arraybackslash}p{1.1cm}@{\hskip 3pt}p{1.1cm}>{\raggedleft\arraybackslash}p{1.1cm}@{\hskip 3pt}p{1.1cm}p{0.8cm}}
    \toprule
       \textbf{Benchmark} & 
       \textbf{Board} & 
       \multicolumn{2}{c}{\textbf{LUTs}} & 
       \multicolumn{2}{c}{\textbf{FFs}} & 
       \multicolumn{2}{c}{\textbf{BRAM}} & 
       \multicolumn{2}{c}{\textbf{DSPs}} & 
       \textbf{Frequency [MHz]} \\
    \midrule
        \multirow{3}{*}{STREAM} & 520N & 176,396 & (25\%) & 449,231 & (25\%) & 4,029 & (34\%) & 128 & (2\%) & 316.67 \\
         & U280 DDR & 20,832 & (1.90\%) & 39,002 & (1.39\%) & 558 & (34.19\%)& 160 & (1.78\%) & 300.00\\ 
          & U280 HBM2 & 331,904 & (20.69\%) & 574,976 & (27.24\%)& 1,408 & (77.70\%) & 2,560 & (28.38\%) & 370.00\\ 
            & PAC SVM & 103,628 & (14.53\%) & 244,354 & (7.42\%) & 74 & (0.66\%)  & 32 & (0.56\%) & 346.00\\
    \midrule
        \multirow{3}{*}{RandomAccess} & 520N & 115,743 & (18\%) & 253,578 &(18\%) & 489 & (4\%)& 14 & ($<1$\%)& 329.17 \\
         & U280 DDR & 7,256 &(0.65\%) & 11,716 &(0.50\%) & 38 &(2.23\%) & 14 &(0.16\%s) & 446.00 \\ 
          & U280 HBM2 & 116,096 &(10.68\%) & 187,456 &(8.76\%) & 608 &(33.55\%) & 224 &(2.48\%) & 450.00\\
        & PAC SVM& 103,397& (12\%) & 225,293 & (12\%) & 535 & (5\%)  & 0 &(0\%) & 322.00 \\
    \bottomrule
    \end{tabular}
    \label{tab:resource_usage}
\end{table*}


These benchmarks contain simple kernels that allow synthesis with the Intel and the Xilinx Vitis toolchain without changes in the host or kernel code.


The used synthesis parameters for all benchmarks are given in Table~\ref{tab:used_parameters}.
The resulting resource usage of the synthesized kernels for STREAM and RandomAccess are given in Table~\ref{tab:resource_usage} and for the remaining benchmarks of the suite in Table~\ref{tab:resource_usage_others}.
A straight forward comparison of the resource usage between the \ac{FPGA} boards is not possible because hardware and software architectures are different.
Nevertheless, in the table a general overview of the resource usage is given by looking at the basic resource elements of an \ac{FPGA}: The \acp{LUT}, \acp{FF}, \ac{BRAM} and \acp{DSP}.
The table only takes into account the resources directly used for the kernels.
Next to the absolute value, the percentage of the used resources relative to the available resources is given.
Absolute values and ratios of the resource usage are taken directly from the reports generated by the \ac{HLS} tools.

\begin{table}
    \centering
    \caption{Measurement Results for STREAM and RandomAccess on different FPGA boards}
    \begin{tabular}{p{2.9cm}>{\raggedleft\arraybackslash}p{1cm}>{\raggedleft\arraybackslash}p{1cm}>{\raggedleft\arraybackslash}p{1cm}>{\raggedleft\arraybackslash}p{1cm}}
    \toprule
        \textbf{Benchmark} & \textbf{520N} & \textbf{U280 HBM2} & \textbf{U280 DDR} & \textbf{PAC SVM}\\
         \midrule
        STREAM Copy [\si{\giga\byte\per\second}] & 67.01 & \rev{377.42} & 33.94 & 20.15\\
        STREAM Scale [\si{\giga\byte\per\second}] & 67.24 & \rev{365.80} & 33.92 & 20.04\\
        STREAM Add [\si{\giga\byte\per\second}] & 68.90 & \rev{374.03} & 34.58 & 15.04\\
        STREAM Triad [\si{\giga\byte\per\second}] & 68.90 & \rev{378.88} & 34.57& 11.66 \\
        STREAM PCIe read [\si{\giga\byte\per\second}] & 6.41 & \rev{6.66} & 5.68 & --\\
        STREAM PCIe write [\si{\giga\byte\per\second}] & 6.32 & \rev{6.03} & 5.47 & --\\
        \midrule
        Random Access Updates [MUOP/s] & 245.0 & \rev{128.1} & 40.3 & 0.5\\
        Random Access Error & 0.0099\% & 0.0106\% & 0.0106\% & 0.0106\% \\
         \bottomrule
    \end{tabular}
    \label{tab:measurement_results_gm}
\end{table}

\subsection{Performance Evaluation}

The synthetic global memory benchmarks STREAM and RandomAccess are executed on three \acp{FPGA}.
The measured performance results are give in Table~\ref{tab:measurement_results_gm}.
The size of the data arrays is set to \rev{$2^{29}$} items, which corresponds to \SI{2}{\giga\byte} of data per data array.
This is the largest power of two that fits into the \SI{8}{\giga\byte} \ac{HBM2} of the Alveo U280 board.

The theoretical peak performance for DDR memory is \SI{19.2}{\giga\byte\per\second} per bank for both the 520N and the U280 boards.
That said, the STREAM benchmark achieves an efficiency between 87.2\% and 90.1\% of the theoretical peak performance for both devices for all four operations.
The local memory buffer allows to use memory bursts to load and store data, which increases the efficiency of  memory accesses.
Nevertheless, the local memory buffer has to be chosen small enough to allow kernel frequencies above the frequency of the memory controller.
The \ac{HBM2} of the U280 board offers a theoretical peak bandwidth of \SI{460}{\giga\byte\per\second}.
\rev{So the benchmark achieves an efficiency of 82.4\% on this board}. 
In contrast to the measurements on 520N and U280, where onboard memory resources DDR or \ac{HBM2} are used, the SVM functionality of the PAC card allows manipulating the array directly in the DDR memory of the host.
The full-duplex PCIe connection makes it possible to define the local buffer very small to allows the implementation of the same in \acp{FF}.
Read and write bursts can still be applied because the kernel will be implemented in a single pipeline.
For the \emph{Copy} and \emph{Scale} operation the kernel achieves more than \SI{20}{\giga\byte\per\second}.

For the Random Access benchmark, a data array of $2^{29}$ items, which is a total of \SI{4}{\giga\byte}, is equally split onto the available memory banks.
All kernels have to calculate all addresses and only update the value if it is placed within the range in the data array they are assigned to.
This leads to a compute-bound implementation for a high number of kernel replications since the pipeline stalls caused by memory accesses per kernel will decrease. Still, the amount of random numbers that have to be generated stays the same.
So the maximum achievable updates per second are limited by the kernel frequency.
The measured performance will show a higher gap to the maximum performance for designs with only a small amount of replications because the random memory accesses will consume a considerable amount of time.
The RandomAccess benchmark shows huge performance differences between the used boards.
One reason for that is the difference in the kernel design that allows the creation of a single pipeline in case of the 520N board using the Intel-specific \emph{ivdep} pragma.
This optimization also slightly decreased the calculation error that is introduced by the buffer.
Nevertheless, for the PAC SVM tests this optimization had to be disabled because with 98\% the error drastically exceeds the allowed range.
For Xilinx there is to our knowledge no optimization flag similar to \emph{ivdep}.

\begin{table*}
    \centering
    \caption{Resource usage of the synthesized Benchmark kernels}
    \begin{tabu}{p{1.4cm}p{1.2cm}>{\raggedleft\arraybackslash}p{1.4cm}@{\hskip 3pt}p{1.1cm}>{\raggedleft\arraybackslash}p{1.1cm}@{\hskip 3pt}p{1.1cm}>{\raggedleft\arraybackslash}p{1.1cm}@{\hskip 3pt}p{0.8cm}>{\raggedleft\arraybackslash}p{0.8cm}@{\hskip 3pt}p{1.1cm}p{1.0cm}}
    \toprule
       \textbf{Benchmark} & 
       \textbf{Board} &
       \multicolumn{2}{c}{\textbf{LUTs}} & 
       \multicolumn{2}{c}{\textbf{Registers}} & 
       \multicolumn{2}{c}{\textbf{BRAM}} & 
       \multicolumn{2}{c}{\textbf{DSPs}} & 
       \textbf{Frequency [MHz]} \\
    \midrule
        b\_eff & 520N & 114,064 &(17\%) & 241,619 & (17\%)& 403 &(3\%)& 0 &(0\%) & 286.67\\
        \midrule
        \multirow{3}{*}{PTRANS} & 520N & 118,885 &(17\%)& 249,516& (17\%)& 2,475& (22\%)& 19& ($<1$\%) & 350.00 \\
        \rowfont{\color{\colorrevi}} & PAC SVM & 116,179 &(14\%)& 249,601 & (14\%)& 2,649& (23\%)& 19& ($<1$\%) & 302.00 \\
        \rowfont{\color{\colorrevi}} & U280 DDR & 15,655 &(1.41\%)& 19,886 & (0.86\%)& 279& (16.54\%)& 40 & (0.44\%) & 300.00 \\
        \midrule
        \multirow{2}{*}{FFT} & 520N & 108,107 &(17\%) & 246,448& (17\%) & 1,026& (9\%)& 312& (5\%) & 366.67 \\
        \rowfont{\color{\colorrevi}} & PAC SVM & 121,535 &(14\%)& 268,981 & (14\%)& 1,064& (9\%)& 312& (5\%) & 327.00 \\
        \midrule
        \multirow{3}{*}{GEMM} & 520N & 136,585 &(20\%)& 353,107 &(20\%)& 1,469& (13\%)& 726& (13\%) & 320.84 \\
        \rowfont{\color{\colorrevi}} & PAC SVM & 139,639 &(17\%)& 351,249 & (17\%)& 1,629& (14\%)& 726& (13\%) & 296.00 \\
        \rowfont{\color{\colorrevi}} & U280 DDR & 154,810 &(14.07\%)& 231,897 & (10.07\%)& 222 & (13.30\%)& 2566 & (28.47\%) & 250.00 \\
        \midrule
        \multirow{2}{*}{LINPACK} & 520N & 203,339 &(32\%) & 667,965 &(32\%) & 3,310 &(28\%) & 786 &(14\%) & 166.25\\
        \rowfont{\color{\colorrevi}} & PAC SVM & 208,603 &(27\%)& 654,653 & (27\%)& 3,453& (29\%)& 786& (14\%) & 276.00 \\
    \bottomrule
    \end{tabu}
    \label{tab:resource_usage_others}
\end{table*}

Table~\ref{tab:measurement_results_520n} contains the results for the remaining benchmarks executed on \rev{a subset of the previously used \ac{FPGA} boards}.
The effective bandwidth of the network that is used as a performance metric by the b\_eff benchmark combines the latency and throughput of a channel into a single metric.
\rev{The sent messages are of the size $2^0,2^1,\dots,2^{20}$ Byte.}
The used optical switch does only add a constant latency to the transmission.
So for the performance model we will use the channel latency of \SI{520}{\nano\second} that is given in the \ac{BSP} documentation for the 520N board and the 19.4 \ac{BSP} \cite{bittware_bsp_doc}.
Also, the channel frequency of \SI{156.25}{\mega\hertz} and the maximum channel width of \SI{256}{\bit} is taken from this document.
We use a ring topology in our implementation, which means every board is connected to two other boards.
With a total of four channels on each \ac{FPGA} this results in two channels per connection and a combined width of \SI{512}{\bit}.
For the transfer of a single message with size $m$ the implementation will need $c_m = \lceil\frac{m}{\SI{512}{\bit}}\rceil$ clock cycles.
This can be used to calculate the transmission time $t_m = \frac{c_m}{\SI{156.25}{\mega\hertz}} + \SI{520}{\nano\second}$ which is the number of clock cycles plus the latency of the channel.
The effective bandwidth can then be calculated with $b\_\mathrm{eff} = \frac{\sum_{m \in L} \frac{m}{t_m}}{21}$.
This results in an effective bandwidth of \SI{8.139}{\giga\byte\per\second} per \ac{FPGA}.
Since there is no congestion on the network, the bandwidth is expected to increase linearly when more \acp{FPGA} are added to the network.
Thus, for 8 \ac{FPGA} the expected effective bandwidth is $8 * \SI{8.14}{\giga\byte\per\second} = \SI{65.11}{\giga\byte\per\second}$.
The synthesized kernel executes the send and receive pipeline sequentially instead of simultaneously for both kernels.
This prevents the utilization of the full-duplex channels and is effectively halving the channel bandwidth.
So the kernel will at most achieve an effective bandwidth of \SI{4.07}{\giga\byte\per\second} per \ac{FPGA}.
Thus, the kernel achieves 96.2\% of its modeled performance but only 48.1\% of the maximum bandwidth of the network.

The GEMM benchmark uses $4096 \times 4096$ matrices for the calculation.
The actual calculation is done on $8 \times 8$ matrices defined by the \texttt{GEMM\_BLOCK} parameter.
So the kernel can initialize the calculation of 1024 floating-point multiplications and additions per clock cycle.
Together with the kernel frequency of \SI{320.84}{\mega\hertz}, this leads to a theoretical kernel peak performance of 328.54~GFLOP/s.
All other latencies that might be introduced by the memory or the calculation can be neglected since the kernel is fully pipelined, and they are hidden by the pipelined execution.
The best execution result of 10 runs achieves 321.59~GFLOP/s, which corresponds to an efficiency of 97.9\% of the theoretical kernel performance.
\rev{The kernel for PAC SVM only clocks with \SI{296}{\mega\hertz} and achieves 241.76~GFLOP/s. 
This corresponds to 79.8\% efficiency of the theoretical kernel performance.
With the normalized performance to a kernel frequency of \SI{100}{\mega\hertz}, another efficiency metric is given in the table for the GEMM benchmark.
The efficiency of the theoretical performance for the U280 board with DDR is similar to the PAC SVM.
}

The PTRANS benchmark transposes an $8192 \times 8192$ matrix and adds the result to a matrix of the same size. 
The global memory interface can load and store 16 values per clock cycle resulting in 16 floating-point operations per clock cycle.
With the maximum clock frequency of the global memory interface of \SI{300}{\mega\hertz}, this results in a theoretical peak performance of 4.8~GFLOP/s.
The benchmark achieves a performance of 3.56~GFLOP/s, which corresponds to 74.2\% efficiency.
The high gap between the theoretical and measured performance is caused by pipeline stalls.
\rev{
Also, for the PAC SVM and U280 DDR, the performance efficiency is low and will be discussed in more detail in Section~\ref{sec:discussion}.
}

The FFT benchmark calculates the 1D-FFT of 4096 single-precision complex floating-point values in a batched manner.
So the benchmark will calculate the FFT for 5000 different data sets.
This allows the kernel to fill the pipeline and hide the latency of the calculation.
Similar to PTRANS, the performance of the kernel is global memory bound.
Eight values are loaded and stored to global memory per clock cycle.
With every value, 12 complex floating-point multiplications are calculated, which corresponds to five single-precision floating-point operations.
This leads to a theoretical kernel peak performance of 144~GFLOP/s.
The measured performance is 116.67~GFLOP/s and thus, 81.0\% of the theoretical performance.
\rev{
Also, on the PAC D5005 with SVM, the FFT benchmark achieves a similar efficiency compared to the STREAM Copy results.
}

\rev{The LINPACK benchmark uses a $4096 \times 4096$ matrix for calculation.
Its synthesized kernel shows a high resource usage and a low kernel frequency.}
This indicates a complex kernel design that increases the difficulty for the compiler to place and route the kernel components on the \ac{FPGA} efficiently.
Some more optimization efforts are necessary to simplify the design and allow higher kernel frequencies which will also increase the performance of the kernel.

\begin{table}
    \centering
    \caption{\rev{Measurement Results for the Remaining Benchmark Applications}}
    \begin{tabu}{p{1.3cm}p{1.3cm}>{\raggedleft\arraybackslash}p{3cm}p{1.6cm}}
    \toprule
        \textbf{Benchmark} & \textbf{Board} & \textbf{Result} & \textbf{Error} \\
         \midrule
        b\_eff & 520N & 31.32~GB/s & -- \\
                \midrule
        \multirow{3}{*}{PTRANS} & 520N & 3.56~GFLOP/s (\SI{42.79}{\giga\byte\per\second}) & 3.81470e-06 \\
        \rowfont{\color{\colorrevi}}& PAC SVM & 0.28~GFLOP/s (\SI{3.36}{\giga\byte\per\second}) & 2.39471e-07\\
        \rowfont{\color{\colorrevi}}& U280 DDR & 0.48~GFLOP/s (\SI{3.67}{\giga\byte\per\second}) & 3.81470e-06\\
                \midrule
        \multirow{2}{*}{FFT} & 520N & 116.67~GFLOP/s (\SI{31.11}{\giga\byte\per\second}) & 3.17324e-01 \\
        \rowfont{\color{\colorrevi}}& PAC SVM & 60.30~GFLOP/s (\SI{16.08}{\giga\byte\per\second}) & 3.17324e-01 \\
                \midrule
        \multirow{2}{*}{GEMM} & 520N & 321.59~GFLOP/s \rev{(\SI{100}{\mega\hertz} norm: 100.23~GFLOP/s)} &  1.54499e-06  \\
        \rowfont{\color{\colorrevi}}& PAC SVM & 241.76~GFLOP/s (\SI{100}{\mega\hertz} norm: 81.68~GFLOP/s) &  2.39471e-07  \\
        \rowfont{\color{\colorrevi}}& U280 DDR & 202.62~GFLOP/s (\SI{100}{\mega\hertz} norm: 81.05~GFLOP/s) & 1.43683e-06 \\
                \midrule
        \multirow{2}{*}{LINPACK} & 520N & 7.51~GFLOP/s & 5.96278e+02 \\
        \rowfont{\color{\colorrevi}}& PAC SVM & 3.46~GFLOP/s & 6.54650e+04 \\
         \bottomrule
    \end{tabu}
    \label{tab:measurement_results_520n}
\end{table}

\section{Further Findings and Investigations}
\label{sec:discussion}

In the previous section, we presented benchmark measurements on different \ac{FPGA} platforms, \acp{BSP} and \acp{SDK} and included comparisons with simple performance models.
Now, we go one step further and discuss how the benchmark suite allows to capture specific issues at the interplay between FPGA targets and tools, with focus on the STREAM benchmark. 


\subsection{\rev{Impact of Memory Access Patterns and Data Layout}}

The results of the STREAM benchmark for the 520N and U280 using DDR show that the design allows to utilize all memory banks to a high degree, also for different FPGA architectures and tools.
Nevertheless, the PAC SVM performance varies a lot among the four operations.
A reason for this is the imbalance in read and write accesses for the \emph{Add} and \emph{Triad} operation.
Two arrays are read, but only one array is written, so only half of the available write bandwidth of the PCIe connection can be utilized, which leads to a performance of \SI{15}{\giga\byte\per\second}.
Still, the \emph{Triad} operation shows a performance significantly below this value.
We were able to reproduce a similar effect for the \emph{Add} operation by changing the allocation order of the three arrays from $A$, $B$, $C$ to $C$, $B$, $A$.
This behavior suggests an issue with the banking of the DDR memory between the arrays that are allocated \rev{in} second and third place.
Separating the two arrays by allocating a fourth array between $B$ and $C$ resolved the measured performance differences.
However, since the placement of the arrays can not be controlled as precisely for SVM as it can be done with \ac{OpenCL} buffers, these effects have to be considered when multiple global memory buffers are used with SVM kernels.

\rev{
The RandomAccess benchmark showed that non-subsequent memory accesses do not perform well in the \ac{SVM} mode.
This also reflects in the low performance for PTRANS and GEMM, which are using strided memory accesses in global memory because of their block-wise calculation of the final result.
The strided memory access leads to a reduction of the length of the memory bursts and lower memory bandwidth.
PTRANS does only achieve 16.7\% of the peak memory bandwidth measured with STREAM.
A lower memory bandwidth also affects the overall performance of the compute-bound GEMM benchmark, since the kernel executes global memory accesses and calculation sequentially.
Because of the heavily decreased memory bandwidth, the read and write overhead to global memory increases and harms the overall kernel performance.
A similar effect can also be observed for GEMM executed on the U280 board, which achieves a comparable performance efficiency.
The performance impact could be reduced by reordering the strided memory on the CPU before writing it to the \ac{FPGA}.
This would allow larger memory bursts and bandwidth similar to the values measured with STREAM and would also increase the performance of the named benchmarks on the boards.
Nevertheless, this would also mean that the CPU has to do a considerable amount of pre-processing, which would exceed the goal of the benchmarks to measure the pure \ac{FPGA} performance.
}

\subsection{Impact of Kernel Frequencies}

An additional synthesis for the 520N board with an increased local buffer size to 16,384 resulted in the resource usage and performance given in Table~\ref{tab:stream_16k_results}.
The high \ac{BRAM} usage decreased the maximum kernel frequency to \SI{280}{\mega\hertz}, which is 93.34\% of the frequency of the memory controller.
At the same time, the benchmark achieves up to 92.15\% of the performance compared to the kernel running with more than \SI{300}{\mega\hertz}, which correlates with the frequency reduction.
This shows that the benchmark is also capable of measuring the performance of the board and tools since the kernel frequency is not least depending on the place and route of the kernel components by the compiler.
Another interesting insight in the measurement results is the difference in the performance of \emph{Copy}/\emph{Scale} and \emph{Add}/\emph{Triad}.
Since the kernel has a lower frequency than the memory controller, the kernel itself becomes the performance bottleneck. 
Execution overheads introduced by switching the pipelines now directly affect the kernel performance and are no longer hidden by latencies of the slower memory controller.

\subsection{\rev{Kernel Scheduling on the U280 with \ac{HBM2}}}

\begin{figure}
    \centering
    \includegraphics[width=0.8\linewidth]{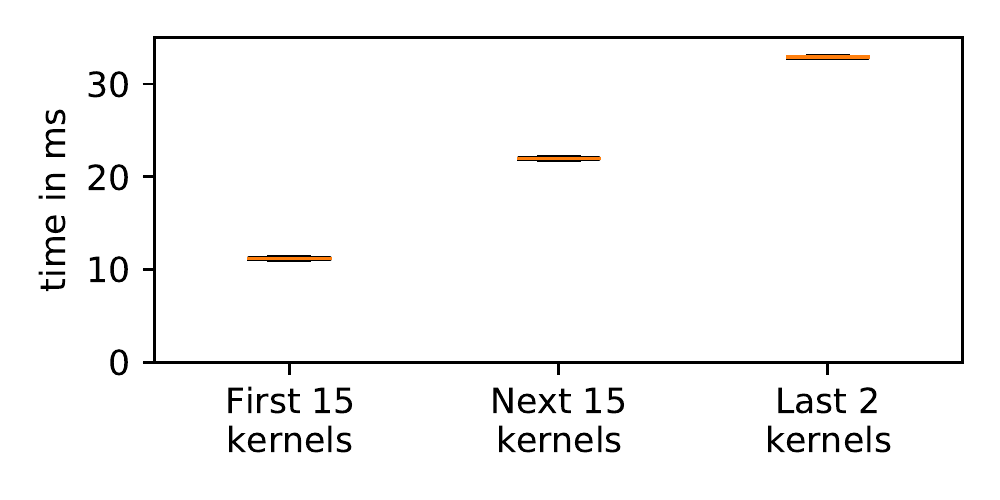}
    \caption{Kernel execution times for the Copy operation on the U280 board with \ac{HBM2} memory using 32 kernel replications. The times are combined in blocks of 15 kernels in the order they where enqueued for execution}
    \label{fig:copy_profiling}
\end{figure}

\rev{In earlier measurements,} the U280 board showed a very low efficiency of below 30\% for all operations of the STREAM benchmark so we further investigated the kernel performance with additional experiments.
The STREAM benchmark measures the time from starting the first kernel until the last kernel has finished execution to calculate the bandwidth.
This is similar to using the slowest runtime of all kernels to calculate the bandwidth.
For a more detailed look \rev{at} the performance of the benchmark, we used the profiling information provided by the \ac{OpenCL} API to collect the execution times separately for every kernel during the benchmark execution.
The used profiling events are the kernel start and end time.
The \rev{thus observed} kernel execution times for the \emph{Copy} operations are given in Figure~\ref{fig:copy_profiling} in groups of 15 kernels with regards to the chronological sequence they were enqueued for execution.
For this measurement, the kernels are enqueued sequentially in two different orders, and the measurements were repeated 20 times.
The second and third group show approximately double and three times the execution time of the first 15 kernels.
\rev{This leads to the hypothesis} that only 15 kernel executions can be maintained simultaneously and that the reported kernel execution time also contains the wait time until the kernel is started.
\rev{
This insight triggered further research into the kernel scheduler of the Xilinx OpenCL runtime system, which actually contains three scheduler implementations. 
After disabling in a configuration file, the two schedulers variants \emph{ERT} and \emph{KDS} that at this time don't support more than 15 concurrent kernels, execution falls back to a scheduler that does not contain these limitations. This mode was used to generate the results in Table~\ref{tab:measurement_results_gm}.
This insight illustrates the importance of a benchmark suite defined on the OpenCL level that measures not only the raw performance potential of FPGA hardware, but also the impact of compilation and synthesis tools and, in this case, runtime environments.
}

\begin{table}
    \centering
    \caption{STREAM benchmark results on the 520N board with 1~MB local memory buffer}
    \begin{tabular}{p{1cm}>{\raggedleft\arraybackslash}p{1.4cm}@{\hskip 3pt}p{.6cm}||p{1.5cm}>{\raggedleft\arraybackslash}p{2cm}}
    \toprule
        \multicolumn{3}{c||}{\textbf{Synthesis}} & \multicolumn{2}{c}{\textbf{Execution}}\\
         \midrule
        \textbf{LUTs} & 203,607 &(26\%) & \textbf{Copy} & \SI{63.48}{\giga\byte\per\second}\\
        \textbf{FFs} & 436,516 &(26\%) & \textbf{Scale} & \SI{63.49}{\giga\byte\per\second}\\
        \textbf{BRAM} & 7,409 &(63\%) & \textbf{Add}   & \SI{58.96}{\giga\byte\per\second}\\
        \textbf{DSPs} & 128 &(2\%) & \textbf{Triad} & \SI{59.00}{\giga\byte\per\second} \\
        \textbf{Freq.} & 280.00 & \si{\mega\hertz} & \textbf{PCIe Write} & \SI{6.40}{\giga\byte\per\second}\\
         &&                    & \textbf{PCIE Read} & \SI{6.32}{\giga\byte\per\second}\\
         \bottomrule
    \end{tabular}
    \label{tab:stream_16k_results}
\end{table}

\subsection{\rev{Power Measurements of the STREAM benchmark on FPGA and CPU}}
\rev{
Another, increasingly important metric in \ac{HPC} is power efficiency.
By default, the presented HPCC FPGA benchmarks do not measure the power consumption since there is no standardized way to retrieve this data.
Nevertheless, we created measurement scripts for each \ac{FPGA} to measure the power consumption of STREAM to investigate the power efficiency of recent \ac{FPGA} boards during execution.
The scripts measured the power consumption during the benchmark execution in intervals of \SI{100}{\milli\second}, and we calculated the average over the whole execution time.
To reduce the FPGA idle time during the measurement, we increased the array sizes to the biggest power of two that fit the device and increased the number of repetitions to 100.
}

\rev{
As a comparison, we executed STREAM v5.10 on the two-socket CPU system equipped with  Intel  Xeon  Gold  6148 that is also used as host for the Nallatch 520N and PAC D5005.
We measured the CPU + DDR power consumption using PCM Tools\footnote{https://github.com/opcm/pcm} (version 202005) in a \SI{1}{\second} interval during execution.
Table~\ref{tab:power_results} presents the power consumption along with the benchmark performance of FPGA targets and CPU reference.
}
\begin{table}
    \centering
    \caption{\rev{Power Consumption vs Benchmark Results for the STREAM benchmark}}
    \begin{tabu}{>{\color{\colorrevi}}p{2.2cm}>{\color{\colorrevi}\raggedleft\arraybackslash}p{.8cm}>{\color{\colorrevi}\raggedleft\arraybackslash}p{.8cm}>{\color{\colorrevi}\raggedleft\arraybackslash}p{.8cm}>{\color{\colorrevi}\raggedleft\arraybackslash}p{.8cm}>{\color{\colorrevi}\raggedleft\arraybackslash}p{.8cm}}
    \toprule
         & \textbf{520N} & \textbf{U280 \ac{HBM2}} & \textbf{U280 DDR} & \textbf{PAC SVM} & \textbf{2x Xeon}\\
         \midrule
        Average [\si{\watt}] & 68.82 & 41.44 & 36.15 & 54.43$^*$ & 323.29 \\
        Peak [\si{\watt}] & 76.78 & 59.98 & 47.09 & 55.21$^*$ &344.91 \\
        Performance per Watt [\si{\giga\byte\per\second\per\watt}] & 0.91 & 6.31 & 0.74 & 0.37$^*$ & 0.53 \\
        \midrule
        Copy [\si{\giga\byte\per\second}] & 69.09 & 372.27& 34.06 & 20.43 & 167.15\\
        Scale [\si{\giga\byte\per\second}] & 69.10 & 374.82 & 34.05  & 20.43& 173.44\\
        Add [\si{\giga\byte\per\second}] & 70.08 & 378.51 & 34.67  & 15.10 & 183.99 \\
        Triad [\si{\giga\byte\per\second}] & 70.02 & 378.55 & 34.66 & 15.12 & 183.40 \\
        PCIe read [\si{\giga\byte\per\second}] & 6.40 & 7.46& 5.44 & -- & -- \\
        PCIe write [\si{\giga\byte\per\second}] & 6.30 & 7.18 & 5.47 & -- & --\\
         \bottomrule
         \multicolumn{6}{l}{\rev{$^*$Additional power consumed by using host DDR memory not included}}\\
    \end{tabu}
    \label{tab:power_results}
\end{table}
\rev{
The power consumption of the benchmark execution is given in the upper part of the table.
The average power consumption is calculated for the whole benchmark execution time.
It needs to be noted that for all \acp{FPGA} executions except for the PAC SVM, this also includes buffer transfer times for every benchmark repetition.
During this time the \ac{FPGA} is nearly idle and consumes less power.
Therefore, the performance per Watt is calculated using the peak measured bandwidth divided by the peak power consumption.
This metric can be used to compare the power efficiency of the devices during the execution of STREAM. FPGA platforms using on-board DDR memory are up to 1.7x more power efficient during this memory-intensive benchmark, the FPGA with \ac{HBM2} is 11.9x more efficient.
}

\subsection{Limitations and Future Work}
\rev{FFT is a memory-bound application that uses subsequent memory accesses to read and write the data to global memory.}
This allows a direct comparison of the performance results to the synthetic STREAM benchmark.
Only two memory banks are used by the kernel \rev{so only half of the total bandwidth can be utilized.
As it can be seen in Table~\ref{tab:measurement_results_520n}, this corresponds to 90.3\% of the \SI{34.45}{\giga\byte\per\second} that are given as upper bound by the STREAM benchmark.
Considering this bandwidth reduction, the FFT benchmark already achieves a high memory utilization.
}
Nevertheless, the application only uses half of the \rev{total} available global memory \rev{banks}.
This could partially be resolved by using memory interleaving or kernel replication.
This also holds for GEMM, where the relatively low resource usage is limiting the kernel performance.
An additional configuration option for replications can be a way to allow matching configurations of the base implementation on different \acp{FPGA} and better utilization of the available resources.

\rev{
The benchmark execution for LINPACK on two different devices showed that the kernel design could not meet the expected performance of this algorithm.
Especially the design of a kernel for the \emph{gefa} routine is a challenging problem because of the complexity of the algorithm.
Previous works show that it is possible to achieve well-performing designs for specific \acp{FPGA} \cite{RodiniaFPGA,fpga_lu_factorization,linpack_fpga}.
Still, these designs are limited in the matrix size or the accuracy because of missing pivoting.
The existing implementations show that the development of a broadly usable base implementation for LINPACK is a challenging task.
Until then, it is also possible and explicitly allowed to execute the benchmark with customized kernels to collect performance results.
}

\section{Conclusion} 
\label{sec:conclusion}
In this paper, we proposed \emph{HPCC FPGA}, a novel \ac{FPGA} \ac{OpenCL} benchmark suite for \ac{HPC}.
Therefore, we provide configurable \ac{OpenCL} base implementations and host codes for all benchmarks of the well-established \ac{HPCC} benchmark suite.
We showed that the configuration options allow the generation of efficient benchmark kernels for Xilinx and Intel \acp{FPGA} using the same source code without manual modification.
We executed the benchmarks on up to three \acp{FPGA} with four different memory setups and compared the results with simple performance models.
Most benchmarks showed a high-performance efficiency when compared to the models.
Nevertheless, the evaluation showed that the base implementations are often unable to utilize the available resources on an \ac{FPGA} board fully.
Hence, it is important to discuss the base implementations and configuration options with the community to create a valuable and widely accepted \ac{FPGA} performance characterization tool for \ac{HPC}.
We made the code open-source and publicly available to simplify and encourage contributions to future versions of the benchmark suite.

\ifopen
\section*{Acknowledgements}

The authors gratefully acknowledge the support of this project by computing time provided by the Paderborn Center for Parallel Computing (PC2). We also thank Xilinx for the donation of an Alveo U280 card, Intel for providing a PAC D5005 loaner board and access to the reference design BSP with SVM support, and the Systems Group at ETH Zurich as well as the Xilinx Adaptive Compute Clusters (XACC) program for access to their Xilinx FPGA evaluation system.

\else
\newpage
\fi

\bibliographystyle{bibliography/IEEEtran}
\bibliography{bibliography/meyer20_sc,bibliography/IEEEabrv}

\end{document}